\documentclass[aps,showpacs,superscriptaddress,twocolumn,amsmath,amssymb,prl]{revtex4-1}
\usepackage{dcolumn}
\usepackage{bm}
\usepackage[dvips]{graphicx}
\usepackage{wrapfig,subfigure}
\usepackage{amssymb}
\usepackage{color}
\usepackage{ulem}

\newcommand{\ph}{\varphi}

\begin{document}

\title{Diffusion-induced bistability of driven nanomechanical resonators}

\author{J. Atalaya}
\affiliation{Department of Applied Physics, Chalmers University of Technology, G{\"o}teborg Sweden, SE-412 96}
\author{A. Isacsson}
\affiliation{Department of Applied Physics, Chalmers University of Technology, G{\"o}teborg Sweden, SE-412 96}
\author{M. I. Dykman}
\affiliation{Department of Physics and Astronomy, Michigan State University, East Lansing, MI 48824}

\begin{abstract}
We study nanomechanical resonators with frequency fluctuations due to diffusion of absorbed particles. The diffusion depends on the vibration amplitude through inertial effect. We find that, if the diffusion coefficient is sufficiently large, the resonator response to periodic driving displays bistability. The lifetime of the coexisting vibrational states scales exponentially with the diffusion coefficient. It also displays a characteristic scaling dependence on the distance to bifurcation points.
\end{abstract}
\date{\today}
\pacs{85.85.+j, 05.40.-a, 62.25.Fg, 68.43.Jk }
\maketitle

Micro and nanomechanical resonators are studied in various contexts, from macroscopic quantum physics \cite{Chan2001a,Blencowe2004a,Kippenberg2008,O'Connell2010} to charge and mass sensing \cite{Steele2009,Lassagne2009,Jensen2008,Naik2009,Lee2010} to atomic and magnetic force microscopy \cite{Rugar2004}. They often have extremely narrow spectral lines at the fundamental mode frequency $\omega_0$, with linewidth $\lesssim 10^{-5}\omega_0$. An important limitation on the linewidth comes from decoherence due to random frequency modulation. In turn, the modulation can itself depend on the vibration amplitude. An example is provided by decoherence due to diffusion of massive particles along the resonator. It occurs because the change of the resonator frequency depends on the particle positions \cite{Naik2009}, but because of inertia the particles themselves are driven toward the antinode of the mode \cite{Lee2010}, as beads on a vibrating string.

\begin{figure}[t]
\includegraphics[width=7.0truecm]{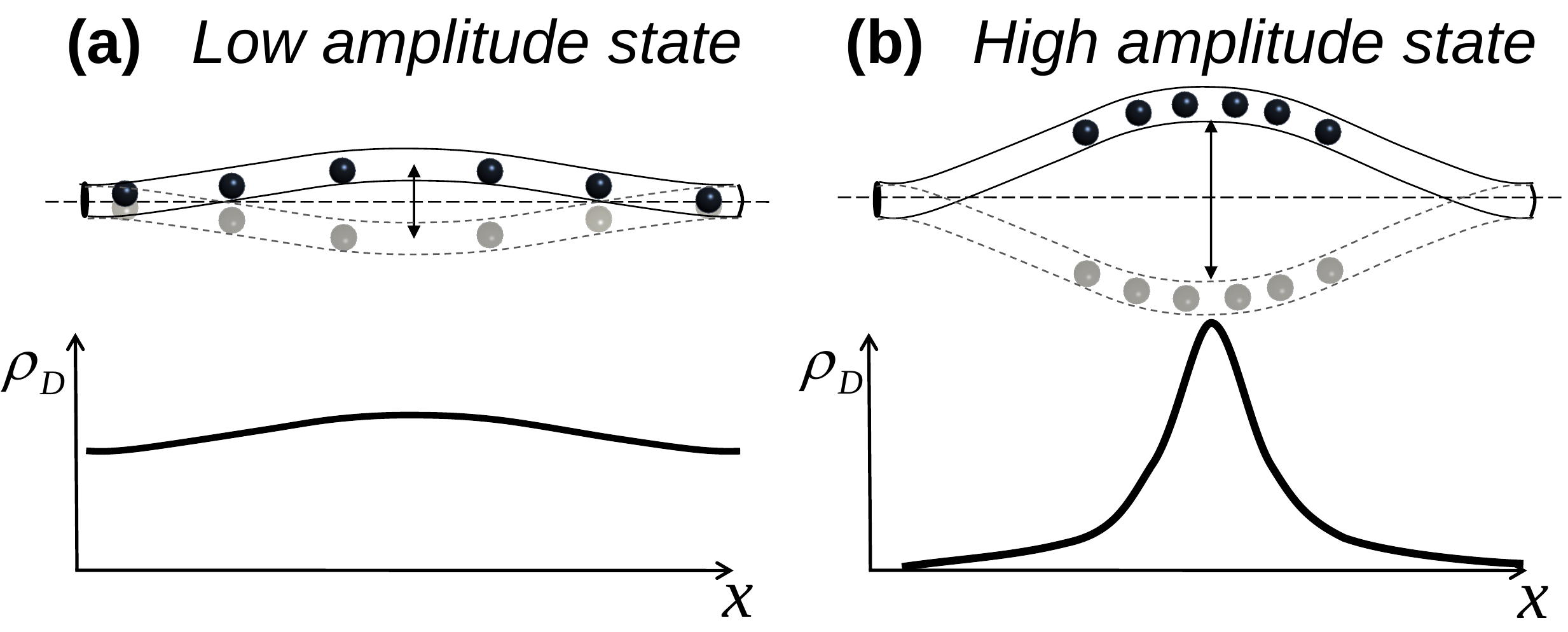}
\caption{Nanobeam resonator with diffusing particles. (a) In the low amplitude vibrational state the particle distribution $\rho_D(x)$ is almost uniform. (b) In the high-amplitude
  state the particles are driven toward the antinode and the distribution has a pronounced maximum there. Because of the
  different mass distributions in (a) and (b), the eigenfrequencies of the resonator are different, which can lead to bistability of resonant response.\label{fig:system} }
\end{figure}

The dynamics of a resonator with diffusing particles becomes particularly interesting in the presence of a comparatively strong periodic driving at frequency $\omega_F$ close to $\omega_0$. Here one can think of the occurrence of two vibrational states. In one of them the vibration amplitude is large, the diffusing particles concentrate close to the mode antinode and tune the vibration frequency of the resonator close to $\omega_F$, leading self-consistently to a large amplitude. In the second state the vibration amplitude is small, diffusion leads to an almost uniform distribution of particles over the resonator, tuning the vibration frequency further away from $\omega_F$, which results, self-consistently, in a small vibration amplitude, see Fig.~\ref{fig:system}.

In this paper we show that driven nanoresonators can display diffusion-induced bistability (DIB), and the DIB may arise even where there are only a few or even one diffusing particle. Besides the physics of nanoresonators, the phenomenon is interesting from a broader perspective. Indeed, diffusion causes frequency fluctuations, and if the system is bistable they would lead to fluctuational switching between the coexisting states $i=1,2$. The notion of bistability is meaningful where the switching rates $W_{ij}$ are much smaller than the relaxation rate $\Gamma$, so that the system stays in one state or the other and only occasionally switches between them. Conventionally, the strong inequality $W_{ij}\ll \Gamma$ requires small noise intensity ${\cal D}$, with $W_{nm}$ displaying an activation dependence on ${\cal D}$ \cite{Kramers1940}. Also, conventionally ${\cal D}$ is proportional to the diffusion coefficient $D$  . As we show, DIB for a few diffusing particles occurs in the opposite limit of large $D$, with $|\ln W_{ij}|\propto D$.

The DIB bears on noise-induced transitions \cite{Horsthemke1984}, where the stationary probability distribution of a system changes from single- to double peaked with increasing noise intensity. The occurrence of such a change does not indicate bistability. The DIB is somewhat closer to the noise-induced bistability seen in numerical simulations of a biomolecular system \cite{Samoilov2005}; however, the conditions for the onset of the bistability and the parameter range where it occurs were not discussed in Ref.~\onlinecite{Samoilov2005}.

The suppression of fluctuations that leads to the DIB for one or a few diffusing particles can be understood for diffusion along a doubly clamped nanobeam, see Fig.~\ref{fig:system}. Such diffusion was seen in recent experiment \cite{Yang2011} and its effect on the spectrum of linear response was analyzed earlier \cite{Atalaya2011}. If the beam length is $L$, for a fixed vibration amplitude the particle distribution is formed over the diffusion time $t_D=L^2/D$ or faster. The distribution depends on the amplitude. In turn, it determines the nanobeam eigenfrequency. Then, if $t_D$ is small compared to the vibration decay time $t_r=1/\Gamma$, one can think of the eigenfrequency as being a function of the instantaneous amplitude. This is a familiar cause of bistability of forced vibrations \cite{LL_Mechanics2004}. Fluctuations are determined by small nonadiabatic corrections, they are generally non-Gaussian, and as we show have intensity $\propto t_D/t_r \ll 1$.

We will consider a nanobeam with $N$ diffusing particles of mass $m$ and assume that $Nm\ll M$, where $M$ is the nanobeam mass. In the fundamental mode, the transverse nanobeam displacement as function of the longitudinal coordinate $x$, see Fig.~\ref{fig:system}, has the form $q(t)\phi(x)$, where $\phi(x)$ describes the spatial profile of the mode and $q(t)$ gives the vibration amplitude; we set $\int \phi^2 dx = L$. The kinetic energy is
\begin{equation}
\label{eq:kin_energy}
{\cal T}_{\rm kin} = \frac{1}{2}\int dx\;\mu(x)\phi^2(x)\dot{q}^2 + \frac{1}{2}\sum\nolimits_{n=1}^{N} m\dot{x}_n^2.
\end{equation}
Here, $x_n$ is the $n$th particle position along the beam and $\mu(x)=M/L+ m \sum\nolimits_n\delta(x-x_n)$. From Eq.~(\ref{eq:kin_energy}), the particles change the nanobeam eigenfrequency from its value without them $\omega_0$ to
$\omega_{\rm e}\approx \omega_0 -(m\omega_0/2M)\sum\nolimits_n \phi^2(x_n).$
At the same time, the vibrations create an effective potential $ -m\dot q^2\phi^2(x)/2$ for the diffusing particles.

In the presence of friction and a force $F\cos\omega_Ft$, the vibrations are described by equation $\ddot q + 2\Gamma\dot q + \omega_{\rm e}^2 q=(F/M)\cos\omega_Ft$. We will assume that $
\Gamma, t_D^{-1}, |\omega_F-\omega_0| \ll \omega_F 
$
and separate fast oscillations at frequency $\omega_F$ from their slowly varying amplitude and phase. To this end, we change from $q,\dot q$ to dimensionless slow complex variables $u,u^*$
\begin{equation}
\label{eq:slow_variable}
u(t)=(2M\Gamma/F)(\omega_F q - i\dot q)\exp(-i\omega_Ft),
\end{equation}
dimensionless slow time $\tau = \Gamma t$, and dimensionless particle coordinates $z_n=x_n/L$. In the neglect of fast-oscillating terms, from Eq.~(\ref{eq:slow_variable})
\begin{eqnarray}
\label{eq:EOM_resonator}
\frac{du}{d\tau} &=& K_r + K_D,\qquad K_r=-\left(1+i\Omega\right)u -i,\nonumber\\
 K_D &=& -iu\sum\nolimits_n \nu(z_n), \qquad
\Omega  = (\omega_F-\omega_0)/\Gamma,
\end{eqnarray}
where $K_r$ and $K_D$ describe the resonator dynamics without diffusing particles and the effect of these particles, respectively;
$\nu(z)=(m\omega_0/2\Gamma M)\phi^2(Lz)$ is the scaled frequency shift of the resonator due to a particle at point $x=Lz$. We disregarded thermal noise in Eq.~(\ref{eq:EOM_resonator}); it is usually weak for nanoresonators and does not change the qualitative results below.

The diffusing particles are usually overdamped. Their equation of motion in the neglect of fast-oscillating terms and particle-particle interaction is
\begin{eqnarray}
\label{eq:EOM_particle}
\frac{dz_n}{d\tau} &=& -|u|^2\partial_{z_n}\Phi(z_n) + \theta^{1/2}\xi(\tau), \quad \theta = l_D^2/L^2.
\end{eqnarray}
Here, $\Phi(z)=-(F/4M\Gamma L)^2(\Gamma\kappa)^{-1}\phi^2(Lz)$, with $\kappa$ being the friction coefficient. The scaled potential $|u|^2\Phi(z)$ is due to beam vibrations. Function $\xi(\tau)$ is white Gaussian noise that leads to diffusion, $\langle \xi(\tau)\xi(\tau')\rangle = 2\delta(\tau - \tau')$.

Parameter $\theta$ in Eq.~(\ref{eq:EOM_particle}) is determined by the diffusion length during the resonator relaxation time $l_D=(D/\Gamma)^{1/2}$ ; $\theta^{-1}\ll 1$ is the large parameter of the theory.  The noise term in Eq.~(\ref{eq:EOM_particle}) is thus not small.

The probability density of the system $\rho=\rho(u,u^*;\{z_n\})$ is given by the Fokker-Planck equation
\begin{eqnarray}
\label{eq:FPE}
\partial_{\tau}\rho&=&-\left\{\partial_u\left[(K_r+K_D)\rho\right] + {\rm c.c.}\right\} +\sum\nolimits_nL_{z_n}\rho, \nonumber\\
L_z\rho &=&\partial_{z}\left[|u|^2\partial_{z}\Phi(z) +\theta\partial_{z}\right]\rho.
\end{eqnarray}
The boundary conditions are $\rho\to 0$ for $|u|\to \infty$ and that, for each particle, the current is zero at the nanobeam boundaries $z=\pm 1/2$ .

We will study first the most interesting case where there is just one particle on the nanobeam. Here, for $\theta\gg 1$ one can use the adiabatic approximation, in which the particle {\it distribution} adjusts to the slowly varying amplitude $|u|$ and phase $\ph=-(i/2)\ln(u/u^*)$ of nanobeam vibrations. We consider an auxiliary eigenvalue problem
\begin{eqnarray}
\label{eq:eigenproblem}
&&L_z\psi_{\alpha} = -\lambda_{\alpha}\psi_{\alpha}
\end{eqnarray}
with boundary conditions $\left[|u|^2\partial_{z}\Phi +\theta\partial_{z}\right]\psi_{\alpha}=0$ for $z=\pm 1/2$, where both $\lambda_{\alpha}$ and $\psi_{\alpha}$ depend parametrically on  $|u|^2$. This is a standard Sturm-Liouville problem. It has an eigenstate with zero eigenvalue,
\begin{eqnarray}
\label{eq:zero_eigen}
\lambda_0=0,\qquad \psi_0(z; |u|^2)=Z^{-1}\exp\left[-|u|^2\Phi(z)/\theta\right],
\end{eqnarray}
whereas the eigenvalues with $\alpha\geq 1$ are positive and large, $\lambda_{\alpha>0}\gtrsim \theta$; in Eq.~(\ref{eq:zero_eigen}) $Z=\int dz \exp\left[-|u|^2\Phi(z)/\theta\right]$. From Eqs.~(\ref{eq:FPE}) and (\ref{eq:eigenproblem}), in short time $\tau \sim \theta^{-1}$ the distribution with respect to $z$ approaches a quasistationary, or adiabatic,  value $\psi_0$.

We seek the overall distribution as $\rho=\sum_{\alpha}p_{\alpha}(u,u^*,\tau)\psi_{\alpha}(z;|u|^2)$. Substituting this expression into Eq.~(\ref{eq:FPE}), multiplying by the left eigenvectors $\bar\psi_{\alpha}$ of operator $L_z$, and integrating over $z$, we obtain
\begin{eqnarray}
\label{eq:p_alpha}
&&\partial_{\tau}p_{\alpha}=-\lambda_{\alpha}p_{\alpha}- \left[\partial_u(K_r p_{\alpha}) + {\rm c.c.}\right]-\sum\nolimits_{\beta}k_{\alpha\beta}
p_{\beta}\nonumber\\
&&\;+\sum\nolimits_{\beta}\nu_{\alpha\beta}\partial_{\ph}p_{\beta}, \qquad\nu_{\alpha\beta}=\int dz \bar\psi_{\alpha}(z)\nu(z)\psi_{\beta}(z).
\end{eqnarray}
Here, $k_{\alpha\beta}=\int dz\bar\psi_{\alpha}(K_r\partial_u+{\rm c.c.})\psi_{\beta}$; we note that, since $\bar\psi_0=1$, $k_{0\alpha}=0$.

From Eq.~(\ref{eq:p_alpha}), function $p_0$ evolves on dimensionless time $\sim 1$. The relaxation time of functions $p_{\alpha}$ with $\alpha > 0$ is $\lambda_{\alpha}^{-1}\lesssim \theta^{-1}\ll 1$. In this time they reach quasistationary values, which can be found from Eq.~(\ref{eq:p_alpha}) by setting $\partial_{\tau}p_{\alpha>0}=0$. Close to a maximum of $p_0$, where $|\partial_{\ph}p_0|\lesssim p_0$, we have $p_{\alpha}\approx -(k_{\alpha 0}p_0 - \nu_{\alpha 0}\partial_{\ph}p_0)/\lambda_{\alpha} \sim p_0/\theta$. To zeroth order in $\theta^{-1}$, in Eq.~(\ref{eq:p_alpha}) for $p_0$ the terms with $p_{\alpha > 0}$ can be disregarded. This leads to the mean-field approximation, $p_0\approx p_{\rm MF}$,
\begin{equation}
\label{eq:pMF}
\partial_{\tau} p_{\rm MF}=-\partial_u(\tilde K_r p_{\rm MF}) + {\rm c.c.},\quad \tilde K_r=K_r - iu\nu_{00}.
\end{equation}
Parameter $\nu_{00}\equiv \nu_{00}(|u|^2)$ gives the shift of the nanoresonator frequency, $\omega_0\to \omega_0-\Gamma\nu_{00}$, which is determined by the vibration amplitude and the diffusion coefficient through Eq.~(\ref{eq:zero_eigen}).

Equation (\ref{eq:pMF}) corresponds to fluctuation-free deterministic motion in the rotating frame $du/d\tau = \tilde K_r$. The stationary states $u=$~const determine periodic vibrational states of the resonator. Their scaled amplitude can be found from equation
\begin{equation}
\label{eq:MF_amplitude}
|u_{\rm st}|^2=\left\{1+\left[\Omega+\nu_{00}(|u_{\rm st}|^2)\right]^2\right\}^{-1}.
\end{equation}
Equation (\ref{eq:MF_amplitude}) can have 1 or 3 solutions. The case of 3 solutions corresponds to the diffusion-induced bistability qualitatively explained in Fig.~\ref{fig:system}; only the solutions with the largest and smallest $|u_{\rm st}|^2$ are stable.

\begin{figure}[t]
\includegraphics[width=8.0truecm]{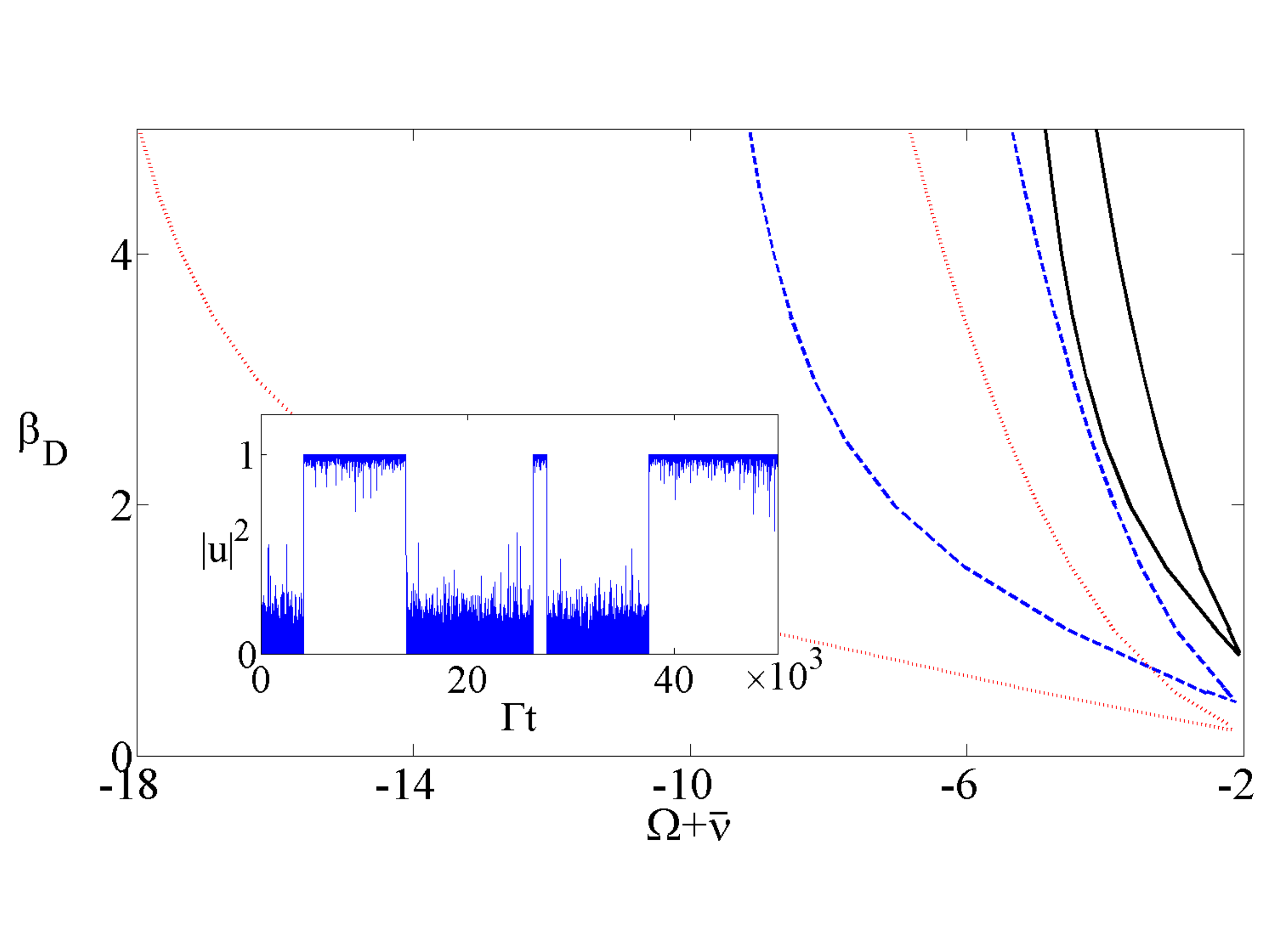}
\caption{Regions of bistability of a nanoresonator on the plane of the scaled intensity $\beta_D= (F/4M\Gamma)^2/\kappa D$ and frequency detuning $\Omega$ of the modulating field. The solid, dashed and dotted lines show the pairs of bifurcation lines for $\bar\nu= Nm\omega_0/2M\Gamma = 5,10,20$, respectively, and for $\phi(x)=2^{1/2}\cos(\pi x/L)$; the bistability occurs inside the corresponding wedges. The inset shows a simulated scaled squared amplitude of forced vibrations $|u|^2$ as function of the scaled time $\Gamma t$ for $\theta=1.01, \bar \nu=10,\Omega=-18.95,\beta_D=5$.
}
\label{fig:bistability_regions}
\end{figure}

The nonlinear response of the resonator displays the dependence on the field amplitude $F$ and frequency $\omega_F$, which is similar to the familiar response of an oscillator with cubic nonlinearity \cite{LL_Mechanics2004}. The parameter range where the bistability occurs is limited by the bifurcation lines where two solutions of Eq.~(\ref{eq:MF_amplitude}) merge. This range has a characteristic wedge-like shape shown in Fig.~\ref{fig:bistability_regions}.

A major effect of fluctuations caused by diffusion and disregarded in Eq.~(\ref{eq:pMF}) is switching between the coexisting vibrational states. Such switching can be seen in the inset of Fig.~\ref{fig:bistability_regions} that illustrates the time dependence of $|u|^2$ obtained by numerical simulations. For small $\theta^{-1}$ the resonator mostly performs small-amplitude fluctuations about the stable states determined by Eq.~(\ref{eq:MF_amplitude}). However, occasionally there occurs a large fluctuation that drives it sufficiently far away to cause switching.

The switching rates are determined by the tail of the probability distribution for $u$ far from its stable values.
The tails of functions $p_{\alpha}$ in Eq.~(\ref{eq:p_alpha}) are generally steep, with $\partial_u p_{\alpha}\propto \theta p_{\alpha}$. Therefore the approximation that led to Eq.~(\ref{eq:pMF}) does not apply on the distribution tail.

The analysis simplifies if the system is close, but not too close to a saddle-node bifurcation point where a stable state merges with an unstable state and disappears \cite{Guckenheimer1987}. Here the rate of switching from the stable state $W$ becomes comparatively large while still $W\ll \Gamma$, which facilitates observing switching; the range is also interesting because $W$ often displays scaling behavior \cite{Dykman1980,Vijay2009,Stambaugh2006}. 

For $\Omega$ (or $F$) close to a bifurcational value, $\Omega_B$ (or $F_B$), the mean-field equation $du/d\tau=\tilde K_r$ is simplified. If we write $u=u_B+u'+iu''$, where $u_B$ is the stationary value of $u$ at the bifurcation point, we find that the relaxation time of $u'$ is 1/2, whereas  $u''$ varies with time slowly for small $|u-u_B|$. By adiabatically eliminating $u'$, we obtain $du''/d\tau \approx \eta-b u''^2$, where $\eta\propto (\Omega-\Omega_B)$ is the distance to the bifurcation point, $|\eta|\ll 1$, whereas $|b|\sim 1$. The bistability exists for $\eta/b > 0$. 

Since $|\tilde K_r|\ll 1$ for small $|\eta|$ and $|u-u_B|$, one can assume, and check afterwards, that in Eq.~(\ref{eq:p_alpha}) $\partial_u p_{\alpha}/\theta p_{\alpha}$ is also small even where $|\partial_up_{\alpha}/p_{\alpha}|\gg 1$. Then the quasistationary solution of Eq.~(\ref{eq:p_alpha}) is $p_{\alpha}\approx \nu_{\alpha0}\partial_{\ph}p_0/\lambda_{\alpha}$ for $\alpha>0$. Substituting this into Eq.~(\ref{eq:p_alpha}) for $p_0$ we obtain
\begin{eqnarray}
\label{eq:p_0}
&&\partial_{\tau}p_0=-\left[\partial_u(\tilde K_r p_0) + {\rm c.c.}\right]+ {\cal D}\partial_{\ph}^2p_0,\\
&&{\cal D}=\sum_{\alpha\geq 1}\frac{|\nu_{0\alpha}|^2}{\lambda_{\alpha}}=\int_0^{\infty}d\tau \langle[\nu\bigl(z(\tau)\bigr) -\nu_{00}]\nu\bigl(z(0)\bigr)\rangle_u.\nonumber
\end{eqnarray}
Here, $z(\tau)$ is given by Eq.~(\ref{eq:EOM_particle}) with $|u|^2=$~const and $\langle\ldots\rangle_u$ means averaging with $|u|^2=$~const. Since we are close to the bifurcation point, ${\cal D}\equiv {\cal D}(|u|^2)= {\cal D}(|u_B|^2)$.

Equation (\ref{eq:p_0}) is a Fokker-Planck equation for the resonator alone. Its form corresponds to $\nu(z)$ in Eq.~(\ref{eq:EOM_resonator}) being white Gaussian noise with mean $\nu_{00}$ and intensity ${\cal D}$. Interestingly, ${\cal D}\propto 1/D$ becomes small for large $D$. This is because $1/D$ gives the correlation time of $\nu(z)$. For slowly varying in time $p_0$ we have ${\cal D}|\partial_u p_0| \sim |\tilde K_r|p_0 \ll p_0$, justifying the earlier assumption.

The analysis of the switching rate near a bifurcation point can be done using the method of Ref.~\onlinecite{Dykman1980}. It gives
\begin{equation}
\label{eq:switch_rate}
W=(\Gamma|\eta b|^{1/2}/\pi)\exp\left(-4|\eta|^{3/2}/3{\cal D}u_B'^2|b|^{1/2}\right),
\end{equation}
where $u_B'={\rm Re}~u_B$. From Eq.~(\ref{eq:switch_rate}), the rate $W$ displays activation dependence on ${\cal D}\propto 1/D$. Also, $\ln W$ scales with the distance to the bifurcation point $\eta$ as $\eta^{3/2}$.

We now consider the case of many diffusing particles, $N\gg 1$; still we assume that their total mass is small, $Nm \ll M$. To the leading order in $1/N$ fluctuations from the particle diffusion are averaged out in the equation of motion for the resonator (\ref{eq:EOM_resonator}). This leads to the mean-field equation for the resonator stationary states $\tilde K_r=0$, where now $\tilde K_r=K_r-iuN\nu_{00}$, i.e., $m$ for the single particle case is replaced by $Nm$.

Fluctuations for $N\gg 1$ are small. For given $Nm$, one can think of the term $\zeta(\tau)=N^{-1}\sum_n \nu \left(z_n(\tau)\right)$ in Eq.~(\ref{eq:EOM_resonator}) as Gaussian noise with intensity $\propto 1/N$. Generally, $\zeta(\tau)$ is not $\delta$-correlated. However, if the relaxation time $t_D$ of diffusing particles is small, $\theta^{-1} \ll 1$, or if the system is close to a bifurcation point, so that its relaxation time is long ($\propto |\eta|^{-1/2}$) and largely exceeds $t_D$, the noise becomes effectively $\delta$-correlated. Its intensity is ${\cal D}/N$, and the distribution of the resonator is described, respectively, by the Fokker-Planck equation (\ref{eq:p_0}) with ${\cal D}$ replaced by ${\cal D}/N$. For large $N$ and $\theta^{-1}\ll 1$ this equation is not limited to the vicinity of the bifurcation point.

\begin{figure}[t]
\includegraphics[width=8.0truecm]{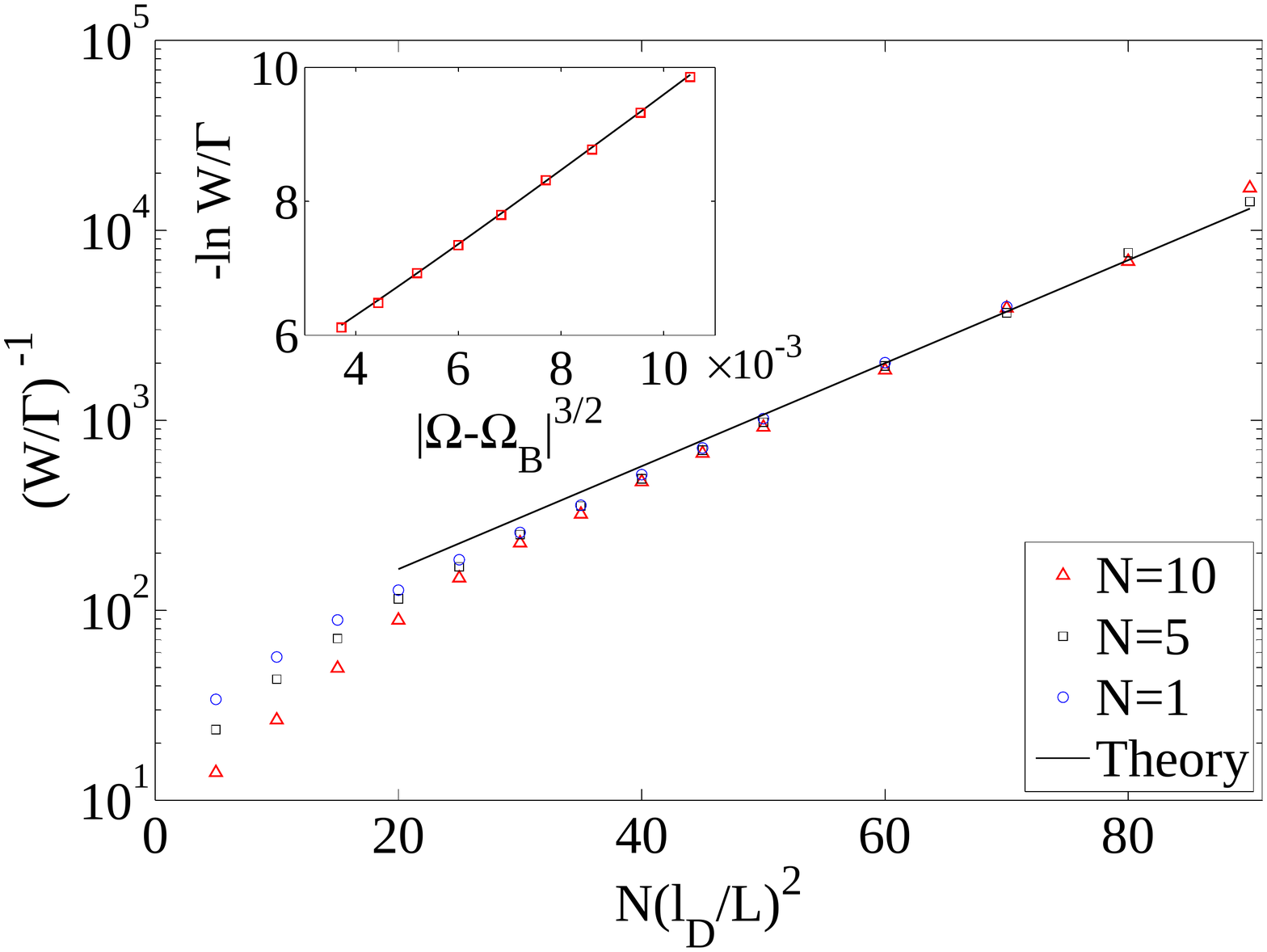}
\caption{The switching rate $W$ near a bifurcation point as function of the product of the diffusion coefficient $D$ and the number of particles $N$ for $\beta_D=5, \bar\nu=10, \Omega-\Omega_B=0.03$ ($\Omega_B=-19.145$). Inset: scaling of $\ln W/ \Gamma$ with the distance to the bifurcation point $\Omega-\Omega_B$ for the same $\beta_D$ and $\bar\nu$; $\theta=5.1$ and $N=1$. The data are the results of simulations, the solid lines show the analytical predictions.}
\label{fig:scaling}
\end{figure}

The above analysis shows that the switching rate should display activation dependence on ${\cal D}/N\propto 1/ND$. This was indeed found in numerical simulations, as seen in Fig.~\ref{fig:scaling}, where $|\ln W|\propto ND$ for large $ND$ and does not depend on $N$ otherwise, for fixed $Nm$. The simulations also demonstrate the 3/2-scaling with the distance to the bifurcation point.

The DIB should arise in nanoresonators and should display the described characteristic behavior  provided the time of diffusion of attached particles over the resonator length exceeds the oscillation period but is smaller than the vibration decay time. For frequency $\omega_0/2\pi= 300$~MHz, decay rate $\Gamma=10^{-5}$~s$^{-1}$, and  length  $L=1~\mu$m the appropriate range of the diffusion coefficient $D$ is $10^{-3}-1~$cm$^2/$s. In addition, the particle mass should not be too small. Plausible candidate systems are provided, for example, by carbon nanotube based nanoresonators with small metallic clusters diffusing along them.

In conclusion, we have demonstrated that diffusion of particles in a nanomechanical resonator can cause bistability of forced vibrations. The bistability can arise even for a single particle, given that the diffusion coefficient $D$ is sufficiently large. In this case the rate of switching scales as $-\ln W\propto D$ and is much smaller than the relaxation rate of the resonator. We also find the scaling behavior of $W$ near bifurcation points. The analytical results are in excellent agreement with simulations, including the scaling of $W$ with the number of particles.     

JA and AI acknowledge the Swedish Research Council and the Swedish Foundation for Strategic Research for the financial support. The research of MID was supported by DARPA and by NSF No. Grant CMMI-0900666.


%

\end{document}